\documentclass[12pt]{article}

\def\slash#1{\setbox0=\hbox{$#1$}#1\hskip-\wd0\hbox to\wd0{\hss\sl/\/\hss}}
\usepackage{epsf, cite}
\usepackage{epsfig}
\usepackage[all]{xy}
\usepackage{amsfonts}
\usepackage{latexsym}
\usepackage{amsmath}
\usepackage{amssymb}

\usepackage{epsf, cite}

\makeatletter
\renewcommand\section{\@startsection {section}{1}{\z@}%
                                   {-3.5ex \@plus -1ex \@minus -.2ex}%nn
                                   {2.3ex \@plus.2ex}%
                                   {\normalfont\large\bfseries}}
\renewcommand\subsection{\@startsection{subsection}{2}{\z@}%
                                     {-3.25ex\@plus -1ex \@minus -.2ex}%
                                     {1.5ex \@plus .2ex}%
                                     {\normalfont\bfseries}}
\makeatother

\let\non\nonumber

\def\lbldef#1#2{\expandafter\gdef\csname #1\endcsname {#2}}

\def\href#1#2{#2}
\def\beq{\begin{equation}}
\def\eeq{\end{equation}}
\def    \bea    {\begin{eqnarray}}
\def    \eea    {\end{eqnarray}}

\renewcommand{\a}{\alpha}
\renewcommand{\b}{\beta}
\renewcommand{\c}{\sigma}
\newcommand{\dl}{\delta}
\newcommand{\e}{\epsilon}
\renewcommand{\r}{\rho}

\def\g{\gamma}

\renewcommand{\H}{{H}}
\def\M{\mathcal G}
\def\L{\mathcal L}
\def\K{\mathcal K}
\def\R{\mathcal R}

\def\S{\mathcal S}
\def\W{\mathcal W}
\def\D{\mathcal D}
\newcommand{\X}{\mathbb{X}}
\renewcommand{\d}{\partial}
\newcommand{\G}[3]{\Gamma^{#1}_{\ #2 #3}}

\providecommand{\openone}{\leavevmode\hbox{\small1\kern-3.8pt\normalsize1}}
\def\xt{\tilde{X}}

\begin{document}
\pagestyle{plain}

\begin{titlepage}

\begin{center}

\hfill{QMUL-PH-2007-25} \\

\vskip 1cm

{{\Large \bf Duality Symmetric Strings, Dilatons \linebreak and O(d,d) Effective Actions}} \\

\vskip 1.25cm {David S. Berman\footnote{email: D.S.Berman@qmul.ac.uk} and
Daniel C. Thompson\footnote{email: D.C.Thompson@qmul.ac.uk}}
\\
{\vskip 0.2cm
Queen Mary College, University of London,\\
Department of Physics,\\
Mile End Road,\\
London, E1 4NS, England\\
}
\end{center}
\vskip 1 cm

\begin{abstract}
\baselineskip=18pt\
We calculate the background field equations for the T-duality symmetric string building on previous work by including the effect of the Dilaton up to two-loops.  Inclusion of the Dilaton allows us to obtain the full beta functionals of the duality symmetric sigma model.  We are able to interpret the result in terms of a dimensionally reduced O(d,d) invariant target space effective action.

\end{abstract}

\end{titlepage}

\pagestyle{plain}

\baselineskip=19pt

\section{Introduction}
Recently there has been much interest in the study of non-geometric backgrounds and in particular T-folds \cite{tfold1}.  These are target spaces in which locally geometric patches are glued with transition functions taking values in the T-duality group.  Although these T-folds are certainly a part of any string landscape they have proven to be surprisingly elusive to construct explicitly.

  To understand string theory more easily on such target spaces it is natural to promote T-duality to be a manifest symmetry of the sigma model action \cite{Dualsym}.   One recent proposal to do just this is the `Doubled Formalism'   in which a target space with a $T^d$ fibration is extended to one with a $T^{2d}$ fibration \cite{hull1, Hull:2006va}.  The additional $d$ bosons in the doubled fibration are constrained so that they are identified with the T-dual.  The sigma-model in this formalism has been shown to be equivalent to the standard sigma-model classically.  Equivalence at the quantum level has been shown for the $d=1$ case in the partition function \cite{bermancopland, Chowdhury:2007ba} and in canonical quantisation with Dirac brackets \cite{emily}.

One important question on the quantum equivalence is to understand the beta functions.  It is not at all obvious that these should be the same as the standard sigma model; they may receive interesting corrections from the T-dual.   This was partially answered in \cite{Berman:2007xn} where the ultra-violet Weyl divergence coefficients of the doubled sigma-model were calculated.  The geometric interpretation of these Weyl divergences was unclear from the point of view of a doubled target space. It was only after integrating out the additional T-dual bosons that the divergences were seen to be equivalent to those obtained for the standard sigma model.

 We should ask what meaning the doubled-beta functions have in their own right without eliminating the dual coordinates.  This is an interesting question since one might hope to find T-fold solutions more readily by trying to find doubled target spaces which satisfy the vanishing of the doubled-beta functions.  However, as we show in this paper, the correct geometric interpretation of the doubled-beta functions is, perhaps surprisingly, not in terms of either the doubled or non-doubled target spaces.  In fact, the beta functions yield the field equations for the dimensional reduction of the standard target space theory.  This dimensionally reduced theory has manifest $O(d,d)$ symmetry. This connection between the manifest $O(d,d)$ theory and T-duality manifest world sheets has been discussed previously in \cite{Maharana:1992my}.

In this letter we shall begin by very briefly reviewing the formalism and approach of \cite{Berman:2007xn}.  We will then extend these results to include the contribution of the Dilaton to first order in $\a^\prime$.  After extracting the full beta functions (related to the Weyl divergence by demanding target space diffeomorphism invariance) we finally demonstrate their target space interpretation.

\subsection{Formalism and Review of Results}
The doubled formalism\cite{tfold1,hull1,Hull:2006va} is an alternate description of string theory on target spaces that are locally $T^d$ bundles, with fibre coordinates $X^i$, over a base $N$ with coordinates $Y^a$.  The fibre is doubled to be a $T^{2d}$ with $2d$ coordinates $\X^A $. The doubled sigma model then has Lagrangian\footnote{The complete formalism also introduces a 1-form connection for the fibration which we set to zero here. There is also a topological term which plays no role in our discussion although it is vital to obtaining the equivalence of the doubled to the non-doubled partition functions \cite{bermancopland}.}
\beq
\mathcal{L} = \frac{1}{4} \H_{AB}(Y)d\X^A\wedge\ast d\X^B  +
\mathcal{L}(Y)
\eeq
where $\mathcal{L}(Y)$ is the standard Lagrangian for a string on the
base and $\H(Y)$ is a metric on the fibre which is assumed to depend only on the base coordinates\footnote{Our conventions are that the worldsheet signature is $(+,-)$, $\d_\pm = \d_0 \pm \d_1$, $\e_{01}= 1 = \e^{01}$ and for convenience we have dropped an overall factor of $2\pi$.  The factor of $\frac{1}{4}$ in (1) is half the usual normalisation and is required to make contact with the standard sigma model.}. One may choose a frame where the metric $\H$ has an $O(d,d)/O(d)\times
O(d)$ coset form as follows:
\bea\label{H}
\H_{AB}(Y) &=& \left( \begin{array}{cc}
h - bh^{-1}b & bh^{-1}\\
-h^{-1}b & h^{-1}
\end{array}\right)\, .
\eea
$h$ and $b$ are the target space metric and two-form field on the fibre of
the un-doubled space. In this frame $\X^A=(X^i,\xt_i)$
with $\{\xt_i\}$ the coordinates on the T-dual torus. To eliminate the extra degrees of freedom, the doubled sigma model is supplemented with a set of worldsheet constraints
\beq
\label{eqConstraint}
d\X^A = L^{AB}H_{BC}\ast d\X^C ,
\eeq
 where the $L$ is an $O(d,d)$ invariant metric such that $\H L^{-1} \H = L$.  In the basis where $\H$ is given by (\ref{H}),
\bea  L_{AB} = \left( \begin{array}{cc}
0& \openone\\
\openone & 0
\end{array}\right).
\eea
We may introduce vielbeins to change to a basis where $\H=\mbox{diag}\left(\openone\,,\openone\right)$, $L=\mbox{diag}\left(\openone\,,-\openone\right)$ and the constraint (\ref{eqConstraint}) is seen to be a chirality constraint on the bosons.  In \cite{Berman:2007xn} this constraint was incorporated into the action using the
method of Pasti, Sorokin and Tonin \cite{Pasti:1996vs}. The resulting action, after gauge fixing the PST symmetry,  looks like two copies of the Floreanini-Jackiw action \cite{FJ}.  In the basis of (\ref{H}) the action we obtain is
\bea\label{Daction}
S= \frac{1}{2} \int d^2\sigma\left[ -\M_{\a\b} \partial_1 X^\a \partial_1 X^\b + \L_{\a\b} \partial_1 X^\a \partial_0 X^\b + \K_{\a\b} \partial_0 X^\a \partial_0 X^\b\right]\,   ,
\eea
where $X^\a = (\X^A, Y^a)$ and
\beq
\M_{\a\b}=
\left(
\begin{array}{cc}
\H_{AB}  &   0 \\
 0 &    g_{ab}
\end{array}
\right),
\L_{\a\b}=
\left(
\begin{array}{cc}
L_{AB} &   0 \\
 0 &    0
\end{array}
\right),
\K_{\a\b}=
\left(
\begin{array}{cc}
0  &   0 \\
 0 &    g_{ab}
\end{array}
\right),
\eeq
in which $g_{ab}$ is the standard sigma model metric for the base\footnote{Notation: Calligraphic objects with Greek indices refer to the total doubled space; uppercase Roman objects with uppercase Roman indices refer to the doubled fibre; lowercase Roman objects correspond to the non-doubled picture with early alphabet lowercase Roman indices denoting the base and late indices denoting the non-doubled fibre.  Hatted objects are those constructed out of the base metric only.}. This action is a generalisation of Tseytlin's duality symmetric formalism \cite{Tseytlin1, Tseytlin2}. The equations of motion on the fibre integrate correctly to give the constraint (\ref{eqConstraint}) and hence the string wave equation.

\subsection{UV Divergences}
In \cite{Berman:2007xn} the ultra-violet divergences in the doubled
formalism were studied by performing a background field expansion \cite{BFM} of the action (\ref{Daction}).  The fields $X^\a$ were written as the sum of a classical piece
$X^\a_{cl}$ and a quantum fluctuation $\xi^\a$.\footnote{Precisely, $\xi$ is  the tangent vector of the geodesic between the classical and quantum values.}  One expands the exponential of the action to quadratic order and takes Wick contractions of $\xi$ to determine the one-loop ultra-violet or Weyl divergence. Since the action (\ref{Daction}) is not manifestly worldsheet Lorentz covariant extra care there are some subtleties and we refer the reader to \cite{Berman:2007xn} for more details.

The result found in \cite{Berman:2007xn} was that the $\L\d_1X\d_0X$ term produced no divergence but that metric term  $\M\d_1X\d_1X$ term did produce divergences with a coefficient which we denote by $\W_{\a\b}$.  These differed from the doubled space Ricci tensor (which is the geometric quantity one would naively have expected) and are given by
\bea
\W_{\a\b} = \R_{\a\b} + \S_{\a\b}\, ,
\eea
where $\R_{\a\b}$ is the doubled space Ricci tensor and
\bea \S_{\a\b} = \left(\begin{array}{cc} 0 & 0 \\ 0 & -\frac{1}{8} \mbox{tr} \left(\d_a\H\d_b\H^{-1}\right)\end{array}\right)\, . \eea
In component form, we have on the base
\bea
\W_{ab}&=&\hat{R}_{ab} + \frac{1}{8}\d_a\H_{AB}\d_b\H^{AB},
\eea
where $\hat{R}_{ab}$ is the Ricci tensor constructed from the base metric $g$ alone.  And on the fibre
\beq
\W_{AB}=-\frac{1}{2}\d^2\H_{AB}+\frac{1}{2}\left((\d_a\H)\H^{-1}(\d^a\H)\right)_{AB}+\frac{1}{2}\G{t}{a}{b}g^{ab}\d_t\H_{AB}\, . \eeq
The divergence differs from the doubled space Ricci tensor by an extra factor of $1/2$ in the term on the base containing the doubled fibre metric $H$.  This is essential to make contact with the standard sigma-model in \cite{Berman:2007xn}, it compensates for a doubled counting coming from doubling the fibre.   This difference however makes it very hard to see a geometric interpretation of the divergence in the doubled picture.  We shall return to address this later but first let us examine how the Dilaton changes the picture.

\section{Inclusion of the Dilaton}
The doubled Dilaton $\Phi$ is related to the standard Dilaton $\phi$ by
\bea
 \Phi(Y) = \phi(Y) + \frac{1}{2}\ln \det h
\eea
and is included into the string action (even in the doubled formalism) with a standard Fradkin-Tseytlin term\footnote{Note the normalisation of $\frac{1}{8\pi}$ which is for later convenience. A different normalisation would require amending the relation between $\Phi$ and $\phi$.}
\bea
\label{Sdil}
S_{dil} = \frac{1}{8\pi}\int d^2\sigma \Phi(Y) R^{(2)}\, .
\eea
We emphasis that this term is an order of $\alpha^\prime$ up on the rest of the doubled string action.  We should remark that in the conformal gauge (which we have used to do the background field expansion) the Dilaton decouples from the theory.

\subsection{Generalised Conformal Invariance}
We can understand how the Dilaton enters into the metric Weyl divergence in a slightly indirect way \cite{Callan:1986jb}.  Consider starting with a sigma model with no Dilaton term.    To demand that the sigma model is Weyl anomaly free we don't actually need that $\W_{\a\b}$ is identically zero.  Since the sigma model should not be affected by on-shell target space field transformation induced by $X^\a \rightarrow X^\a + V^\a$ we only need that the Weyl divergence is zero up to such redefinition.
 Therefor we demand that the full beta function vanishes, that is:
\bea
\b_{\a\b} = \W_{\a\b} - \D_{\a} V_{\b} = 0 \, .
\eea
As explained in \cite{Callan:1986jb} we should restrict $ V_{\b}$ to be the derivative of a scalar which we are led to associate with the Dilaton. Hence
\bea
\b_{\a\b} = \R_{\a\b} + \S_{\a\b} - \D_\a \D_\b \Phi = 0\, .
\eea
We express this on the base and fibre and obtain:
\bea
\label{basebeta}
\b_{ab}&=&\hat{R}_{ab} + \frac{1}{8}\d_a\H_{AB}\d_b\H^{AB} -  \hat{\nabla}_a \hat{\nabla}_b \Phi \,  , \\
\label{fibrebeta}
\b_{AB}&=&-\frac{1}{2}\hat{\nabla}^2\H_{AB}+\frac{1}{2}\left((\hat{\nabla}_a\H)\H^{-1}(\hat{\nabla}^a\H)\right)_{AB}- \frac{1}{2} \hat{\nabla}_a\H_{AB}  \hat{\nabla}^a \Phi\, .
\eea
We have chosen to express these with hatted covariant derivatives constructed out of the base metric $g_{ab}$ and the reader should bear in mind that these objects are actually blind to indices on the fibre e.g.  \[\hat{\nabla}_a\H_{AB} = \d_a \H_{AB}\, , \hat{\nabla}^2\H_{AB}= \d_a\d^a\H_{AB} +\hat{\Gamma}^{a}_{ab}\d^b \H_{AB}\, .\]
 With this notation we are tempted to think of fibre indices as not really labeling target space coordinates but labeling moduli fields contained in $\H$.  In other words it seems like these beta-functionals may have a geometric interpretation in a target space that has been dimensionally reduced so that only the base coordinates remain.  We will see later that this is indeed a sensible interpretation.

\subsection{Dilaton Beta Function}
 Now let us consider the Dilaton beta function.  The leading order part will be the familiar
\bea
\b^\Phi = \frac{26-D}{6} + O(\alpha^\prime)
\eea
where $D$ is the dimension of the non-doubled target space.  The reason for this is seen quite clearly by counting the central charge $c$ of the doubled theory; if we double $d$ coordinates we have $D+d$ bosons however $2d$ of them are chiral and only count half in the central charge.  We thus expect the central charge to remain equal to the standard target space dimension.  The factor of 26 arises from the determinant that arise from integration over world sheet metrics \cite{Polyakov1981}. From here on we shall work in the critical dimension.

The next order in $\alpha^\prime$  contribution is more complicated.  There are
two sources of contributions, namely those which are 1-loop and arise from the background field expansion of $S_{dil}$ to quadratic order and those which are 2-loops and arise from the expansion of the action (\ref{Daction}).

Let us first evaluate the one loop contribution.  Since the Dilaton action is the same as for the standard string we can read off the 1-loop result that
\bea
\b^\Phi &=& \frac{1}{2} \alpha^\prime \left( -2\D_\alpha \D^\alpha \Phi - (D\Phi)^2 +  \mbox{ 2-loops}\right)\\
&=& \frac{1}{2} \alpha^\prime \left( -2\hat{\nabla}_a \hat{\nabla}^a \Phi - (\hat{\nabla}\Phi)^2 +  \mbox{ 2-loops}\right) \,.
\eea
 We have used that $\Gamma^I_{Ia} = \mbox{tr} \H^{-1}\d_{a} \H  = 0$ to get to the second line.

To evaluate the two loop contribution we need to background field expand  (\ref{Daction}) to third and fourth order.  The relevant terms are those that do not directly couple to the classical fields $\d X^\a$ and whose contractions have a power counting of $p^{-2}$.  At third order we find
\bea
S^{(3)} = \frac{1}{2} \int \L_{\a\b;\g} \xi^\g D_0\xi^\a D_1\xi^\b +\K_{\a\b;\g} \xi^\g D_0\xi^\a D_0\xi^\b  + \dots\, ,
\eea
and at fourth order
\bea
\non  S^{(4)} &=&-\frac{1}{6} \int  \R_{\a\c\dl\b} D_1\xi^\a D_1 \xi^\b \xi^\c \xi^\dl\\ \non  && +\frac{1}{48}\int \left[ 12\L_{\a\b;\c\dl} + 4\left( \L_{\a\r}\R^\r_{\, \c\dl \b} +   \L_{\b\r}\R^\r_{\, \c\dl \a} \right)\right] D_0\xi^\a D_1\xi^\b\xi^\c \xi^\dl \\
&& +\frac{1}{48}\int \left[12\K_{\a\b;\c\dl} + 8\K_{\a\r}\R^\r_{\, \c\dl \b}   \right] D_0\xi^\a D_0\xi^\b\xi^\c \xi^\dl + \dots\, ,
\eea
where the dots signify contributions not relevant for the Dilaton.

So to calculate the quantum effective action we will need to calculate the wick contractions of the exponential of the action.  That is, we will have to consider $\langle (S^{(3)})^2\rangle$ and $\langle S^{(4)}\rangle$. One could find the correct propagator contractions by considerations similar to those in the appendix of \cite{Berman:2007xn}.  It would, of course, be a matter of some considerable detail to calculate all the diagrams needed.  However we can quickly see by taking a few test contractions of $\R_{\a\b\c\dl}$ with $\M^{\a\b}, \L^{\a\b}$ and $\K^{\a\b}$ that the result, which must be a scalar, will be given in terms of two underlying objects; the base Ricci scalar $\hat{R}$ and two-derivative contractions of the fibre coset metric $\H$.

So with constants $A$ and $B$ undetermined we will have
\bea
\b^\Phi = \frac{ \alpha^\prime}{2} \left( -2\hat{\nabla}_a \hat{\nabla}^a \Phi -  (\hat{\nabla}\Phi)^2 + A \hat{R} + B \left(\hat{\nabla}_a \H^{-1} \hat{\nabla}^a \H \right)\right)\, .
\eea

\subsection{Integrability Condition}
We shall use an indirect way to determine these coefficients. For the standard string the vanishing of the metric beta function and the Bianchi identity imply that the divergence of the metric beta function is equal to the gradient of some scalar.  This scalar is identified as the Dilaton beta function. We show that with a little extra work the same is true for the doubled case, that is
\bea
\label{intcond}
\D^\a \b(\M)_{\a\b} = \D_\b \b(\Phi)
\eea
It is straight forward to see that
\bea
\D^\a \b(\M)_{\a\b} &=& \D^\a \left( \R_{\a\b} +\S_{\a\b} - \D_\b \D_\a \Phi\right)\\
&=& \frac{1}{2} \D_\beta \left(  \R - 2 \D^2 \Phi - (\D \Phi)^2 \right) + \S_{\a\b}\D^\a \Phi + \D^\a \S_{\a\b}.
\eea
It is also easy to see that on the fibre (when $\b=B$) both sides of (\ref{intcond}) are zero by the assumption that no fields depend on the fibre coordinates. So we will need to evaluate this expression on the base (when $\b=b$).  We find
\bea
\label{intcond2}
\non \D^\a \b_{\a\b} \bigg|_{\b =b} &=&  \hat{\nabla}_b \left( \frac{\hat{R} }{2}  + \frac{1}{8} \mbox{tr} (\hat{\nabla}_a \H \hat{\nabla}^a \H^{-1})   - \hat{\nabla}^2 \Phi - \frac{1}{2}(\hat{\nabla}\Phi)^2 \right)\\ &&  + \left( D^\a \S_{\a\b} +  S_{\a\b}D^\a \Phi \right)\bigg|_{\b =b}.
\eea
To proceed we are going to need a suitable Bianchi-like identity for $\S_{\a\b}$ to allow us to pull out a total derivative. We can find such an expression by consider the base components of the doubled space Bianchi identity,
\bea
D^\a \R_{\a\b} \bigg|_{\b =b} = \frac{1}{2}D_\b \R \bigg|_{\b =b}\, .
\eea
On the right hand side one finds
\bea
\frac{1}{2}\hat{\nabla}_b \R = \hat{\nabla}_b\left( \frac{1}{2}\hat{R} +\frac{1}{8} \mbox{tr} (\hat{\nabla}_a \H^{-1} \hat{\nabla}^a \H )   \right)\, ,
\eea
whilst on the left one finds with a bit of manipulation
\bea
\non D^\a \R_{\a\b} \bigg|_{\b =b} &=& \hat{\nabla}^a\hat{R}_{ab} +  \frac{1}{4} \mbox{tr} (\hat{\nabla}_a \H \H^{-1} \hat{\nabla}_b \H  \H^{-1} \hat{\nabla}^a \H  \H^{-1})  \\ && -  \frac{1}{4}\mbox{tr} (\hat{\nabla}^a \H \H^{-1} \hat{\nabla}_a\hat{\nabla}_b \H  \H^{-1})\,.
\eea
After equating these last two expression (and using the Bianchi identity of the base Ricci tensor) we have an identity that allow us to see that
\bea
\label{intcond3}
\non \frac{1}{2}\hat{\nabla}_b \mbox{tr} (\hat{\nabla}_a \H^{-1} \hat{\nabla}^a \H ) &=& \mbox{tr} (\hat{\nabla}_a \H \H^{-1} \hat{\nabla}_b \H  \H^{-1} \hat{\nabla}^a \H  \H^{-1}) \\ &&- \mbox{tr} (\hat{\nabla}^a \H \H^{-1} \hat{\nabla}_a\hat{\nabla}_b \H  \H^{-1})\,.
\eea

We now look at the $\S$ terms in (\ref{intcond2}). For the first we have
\bea
D^\a \S_{\a\b} \bigg|_{\b =b} &=& -\frac{1}{8} \hat{\nabla}^a  \mbox{tr} (\hat{\nabla}_a \H \hat{\nabla}_b \H^{-1})\\
 \non &=&  -\frac{1}{8}   \mbox{tr} (\hat{\nabla}^2 \H \hat{\nabla}_b \H^{-1}) - \frac{1}{4}   \mbox{tr} (\hat{\nabla}_a \H  \H^{-1} \hat{\nabla}_b \H  \H^{-1} \hat{\nabla}_a \H\H^{-1})
 \\ &&+ \frac{1}{8}   \mbox{tr} (\hat{\nabla}^a \H  \H^{-1} \hat{\nabla}_a \hat{\nabla}_b \H  \H^{-1})\,.
\eea
For the second we find
\bea
\S_{\a\b}D^\a \Phi  \bigg|_{\b=b} = -\frac{1}{8}  \mbox{tr} (\hat{\nabla}_b \H^{-1} \hat{\nabla}_a \H )\hat{\nabla}^a \Phi = -\frac{1}{8}(\hat{\nabla}_b \H^{IJ})\hat{\nabla}_a \H_{IJ}\hat{\nabla}^a \Phi\, .
\eea
We can now use the vanishing of the fibre components of the doubled metric beta function (\ref{fibrebeta}) to swap the Dilaton for some more terms involving $\H$.  We find
\bea
\S_{\a\b} D^\a \Phi  \bigg|_{\b = b}
&=& \frac{1}{8}  \mbox{tr} (\hat{\nabla}_b \H^{-1} \hat{\nabla}^2 \H) + \frac{1}{8} \mbox{tr} (\hat{\nabla}_a \H  \H^{-1} \hat{\nabla}_b \H  \H^{-1} \hat{\nabla}_a \H \H^{-1}) \, .
\eea
Upon invoking the identity (\ref{intcond3}) we find that
\bea
\left( D^\a \S_{\a\b} +  S_{\a\b}D^\a \Phi \right) \bigg|_{\b =b} =  -\frac{1}{16} \hat{\nabla}^b \mbox{tr} (\hat{\nabla}_a \H \hat{\nabla}^a \H^{-1}).
\eea
Thus from the integrability condition (\ref{intcond}) we determine the Dilaton beta function to be
\bea
\label{dilbeta}
\b^\Phi = \frac{ \alpha^\prime}{2} \left( -2\hat{\nabla}_a \hat{\nabla}^a \Phi- (\hat{\nabla}\Phi)^2 + \hat{R} + \frac{1}{8} \left(\hat{\nabla}_a \H^{-1} \hat{\nabla}^a \H \right)\right)\, .
\eea
\section{Target space interpretation}
For the standard string a result of fundamental importance is that the beta functions can be connected to the field equations of a target space theory \cite{Callan:1985ia}.  The identification is schematically given by
\bea
\label{betaid1}\b^G - \frac{1}{\a^\prime}G\b^\phi &\sim& \dl_GS \, ,\\
\label{betaid2}\frac{1}{\a^\prime}\b^\phi &\sim& \dl_\phi S \, ,\\
\label{betaid3} \b^{matter} &\sim& \dl_{matter}  S \, ,
\eea
where $S$ is the action for the target space theory. Given that we have seen that the beta functions appear geometric in a dimensionally reduced target space it is naturally to guess that the correct target space theory is the dimensional reduction of the standard bosonic string target space theory. We now show that this is indeed the case.

Starting with the d=26 bosonic space time action,
\bea
S_{26}= \int d^{26}x \sqrt{-G}e^{\phi}\bigl\{ R(G) + \left(\nabla \phi \right)^2  - \frac{1}{12}H^2\bigr\}
\eea
 we reduce on $T^d$ with the relevant ansatz (in this case a simplified diagonal reduction) given by
\bea
G=
\left(
\begin{array}{cc}
h(Y)  &   0 \\
 0 &    g(Y)
\end{array}
\right) &
B=
\left(
\begin{array}{cc}
b(Y)  &   0 \\
 0 &    0
\end{array}
\right) &
 \phi(Y) = \Phi(Y) - \frac{1}{2}\ln \det h\,.
\eea
With this ansatz we see that the T-dual invariant (i.e. doubled) Dilaton $\Phi$ emerges in the reduced action since
\bea
\sqrt{-G}e^\phi = \sqrt{-g}\sqrt{h}e^\phi = \sqrt{-g}e^{\phi + \frac{1}{2}\ln \det h} = \sqrt{-g}e^\Phi\, .
\eea
This expression also helps understand the choice of normalisation of (\ref{Sdil}).   The standard result is a low energy effective action which displays manifest $O(d,d)$ symmetry e.g. \cite{MyersOdd, Maharana:1992my}
\bea
S_{26-d}=  vol(\mathcal{M}^d) \int d^{26-d}y \sqrt{-g} e^{\Phi}\bigl\{ \hat{R}(g)  + \left(\hat{\nabla} \Phi\right)^2
 + \frac{1}{8}\mbox{tr} \left(L \hat{\nabla}_a \H L \hat{\nabla}^a \H\right) \bigr\}\, .
\eea
The fields arising from the internal components of the metric and B-field are thought of as moduli in that they parametrise the vacuum of the dimensionally reduced theory.  They have been organised in an $O(d,d)$ invariant way by placing them in coset metric $\H_{IJ}$.

We can see that the reduced action doesn't `remember' which T-dual compactifaction it arose from.  In this sense it is natural to expect a clear linkage with the doubled formalism.  In performing this dimensional reduction we assumed that the fields had no dependence on the internal coordinates.  This assumption doesn't contradict the general aim of understanding T-folds since a non-trivial fibration over the base is still allowed.

From the variation with respect to the metric we find Einstein's equation:
\bea
 \label{baseeqm} 0 &=& \hat{R}_{ab} - \frac{1}{2}g_{ab}\hat{R} -  T_{ab}\, ,
\eea
where the stress-energy is given as
\bea
\non T_{ab} &=& \hat{\nabla}_a \hat{\nabla}_b\Phi - g_{ab}\hat{\nabla}^2 \Phi -\frac{1}{2} g_{ab}\left(\hat{\nabla} \Phi\right)^2 \\ && - \frac{1}{8}\left( tr \left(\hat{\nabla}_a \H^{-1} \hat{\nabla}_b \H\right)  - \frac{1}{2}g_{ab} tr \left(\hat{\nabla}_a \H^{-1} \hat{\nabla}^a \H\right)\right)\, .
\eea
Varying with respect to the  fields $\H$
\bea
\label{modulieqm}
0 &=& \hat{\nabla}\left( e^\Phi \H^{-1}\hat{\nabla}\H \right)\\
&=& e^\Phi \H^{IJ} \left(\hat{\nabla}^2 \H_{IJ} + \hat{\nabla}_a\Phi\hat{\nabla}^a \H_{IJ}- \left(\hat{\nabla}_a \H \H^{-1} \hat{\nabla}^a \H\right)_{IJ}\right)\, ,
\eea
and from the Dilaton
\bea
\label{dilatoneqm}
0=-2\hat{\nabla}^2\Phi - \left(\hat{\nabla}\Phi\right)^2 +\hat{R} +\frac{1}{8}  \mbox{tr} \left(\hat{\nabla} \H^{-1} \hat{\nabla}\H\right)\, .
\eea

Immediately we see that fibre components of the doubled metric beta function $\b_{IJ}$, given by (\ref{fibrebeta}), are proportional to the field equations for $\H_{IJ}$ (\ref{modulieqm}), the doubled Dilaton beta function $\b^{\Phi}$ is proportional to the field equation for $\Phi$ (\ref{dilatoneqm}) and the identification rules  (\ref{betaid1}, \ref{betaid2} ,\ref{betaid3}) are satisfied with the base beta function $\b_{ab}$ playing the role of the $\b^G$ in (\ref{betaid1}) .

\section{Conclusions}

We conclude by making a few remarks about potential generalisations.  In general the dimensional reduction above includes a $U(1)^{2d}$ gauge field coming from isometries of the $T^d$ and transformations of the B-field on the fibre.  It would be nice to shown that these fields arise when one considers a connection in the doubled-fibration.  However, in this case it becomes more subtle to find the constrained action akin to (\ref{Daction}).  It seems that the PST procedure would result in off-diagonal elements in both $\M$ and $\L$.  This would significantly increase the computational burden in calculating the Weyl divergences.

To summarise, we have shown that the full beta-functions of the T-duality symmetric string led to a O(d,d) invariant dimensionally reduced target space theory. This was perhaps to be expected though the utility of this paper is that the detailed relationship between the string in the doubled geometry and the equations of motion of the background are now made manifest. One hopes this may ultimately lead to a better understanding of how T-folds may work.

\noindent {\bf{Note Added}}
When this paper was in preparation two papers appeared concerning interesting aspects of T-folds and doubled geometry \cite{new}.

\section{Acknowledgements}

This work is in part supported by the EC Marie Curie
Research Training Network, MRTN-CT-2004-512194. DCT is supported by a
STFC studentship. We would like to thank Neil Copland for discussions and J. Maharana for pointing out a key important reference.


\begin{thebibliography}{99}


\bibitem{tfold1}
  C.~M.~Hull,
  ``Global aspects of T-duality, gauged sigma models and T-folds,''
  arXiv:hep-th/0604178.
  %%CITATION = HEP-TH 0604178;%%

  A.~Dabholkar and C.~Hull,
  ``Generalised T-duality and non-geometric backgrounds,''
  JHEP {\bf 0605} (2006) 009
  [arXiv:hep-th/0512005].
  %%CITATION = HEP-TH 0512005;%%

  J.~Shelton, W.~Taylor and B.~Wecht,
  ``Nongeometric flux compactifications,''
  JHEP {\bf 0510} (2005) 085
  [arXiv:hep-th/0508133].
  %%CITATION = HEP-TH 0508133;%%

  J.~Shelton, W.~Taylor and B.~Wecht,
  ``Generalized flux vacua,''
  arXiv:hep-th/0607015.
  %%CITATION = HEP-TH 0607015;%%

  C.~M.~Hull and R.~A.~Reid-Edwards,
  ``Flux compactifications of string theory on twisted tori,''
  arXiv:hep-th/0503114.
  %%CITATION = HEP-TH 0503114;%%

  K.~Becker, M.~Becker, C.~Vafa and J.~Walcher,
  ``Moduli stabilization in non-geometric backgrounds,''
  arXiv:hep-th/0611001.
  %%CITATION = HEP-TH 0611001;%%

\bibitem{Dualsym}

 M.~J.~Duff,
 ``Duality Rotations In String Theory,''
 Nucl.\ Phys.\  B {\bf 335} (1990) 610.
  %%CITATION = NUPHA,B335,610;%%
J.~H.~Schwarz and A.~Sen,
  ``Duality symmetric actions,''
  Nucl.\ Phys.\  B {\bf 411} (1994) 35
  [arXiv:hep-th/9304154].
  %%CITATION = NUPHA,B411,35;%%



\bibitem{hull1}
  C.~M.~Hull,
  ``A geometry for non-geometric string backgrounds,''
  JHEP {\bf 0510} (2005) 065
  [arXiv:hep-th/0406102].
  %%CITATION = HEP-TH 0406102;%%

%\cite{Hull:2006va}
\bibitem{Hull:2006va}
  C.~M.~Hull,
  ``Doubled geometry and T-folds,''
  arXiv:hep-th/0605149.
  %%CITATION = HEP-TH 0605149;%%

\bibitem{Tseytlin1}
  A.~A.~Tseytlin,
  ``Duality Symmetric Formulation Of String World Sheet Dynamics,''
  Phys.\ Lett.\ B {\bf 242} (1990) 163.
  %%CITATION = PHLTA,B242,163;%%

%\cite{Tseytlin:1990va}
\bibitem{Tseytlin2}
  A.~A.~Tseytlin,
  ``Duality Symmetric Closed String Theory And Interacting Chiral Scalars,''
  Nucl.\ Phys.\  B {\bf 350} (1991) 395.
  %%CITATION = NUPHA,B350,395;%%


\bibitem{bermancopland}
  D.~S.~Berman and N.~B.~Copland,
  ``The string partition function in Hull's doubled formalism,''
  Phys.\ Lett.\  B {\bf 649} (2007) 325
  [arXiv:hep-th/0701080].
  %%CITATION = PHLTA,B649,325;%%



  %\cite{Hackett-Jones:2006bp}
\bibitem{emily}
  E.~Hackett-Jones and G.~Moutsopoulos,
  ``Quantum mechanics of the doubled torus,''
  JHEP {\bf 0610} (2006) 062
  [arXiv:hep-th/0605114].
  %%CITATION = HEP-TH 0605114;%%

%\cite{Chowdhury:2007ba}
\bibitem{Chowdhury:2007ba}
  S.~P.~Chowdhury,
  ``Superstring partition functions in the doubled formalism,''
  arXiv:0707.3549 [hep-th].
  %%CITATION = ARXIV:0707.3549;%%

%\cite{Berman:2007xn}
\bibitem{Berman:2007xn}
  D.~S.~Berman, N.~B.~Copland and D.~C.~Thompson,
  %``Background Field Equations for the Duality Symmetric String,''
  arXiv:0708.2267 [hep-th].
  %%CITATION = ARXIV:0708.2267;%%

%\cite{Maharana:1992my}
\bibitem{Maharana:1992my}
  J.~Maharana and J.~H.~Schwarz,
  ``Noncompact symmetries in string theory,''
  Nucl.\ Phys.\  B {\bf 390}, 3 (1993)
  [arXiv:hep-th/9207016].
  %%CITATION = NUPHA,B390,3;%%



%\cite{Pasti:1996vs}
\bibitem{Pasti:1996vs}
  P.~Pasti, D.~P.~Sorokin and M.~Tonin,
  ``On Lorentz invariant actions for chiral p-forms,''
  Phys.\ Rev.\  D {\bf 55} (1997) 6292
  [arXiv:hep-th/9611100].
  %%CITATION = PHRVA,D55,6292;%%

%\cite{Floreanini:1987as}
\bibitem{FJ}
  R.~Floreanini and R.~Jackiw,
  ``Selfdual Fields As Charge Density Solitons,''
  Phys.\ Rev.\ Lett.\  {\bf 59} (1987) 1873.
  %%CITATION = PRLTA,59,1873;%%

%\cite{Honerkamp:1971sh}
\bibitem{BFM}
  J.~Honerkamp,
  ``Chiral multiloops,''
  Nucl.\ Phys.\  B {\bf 36} (1972) 130.
  %%CITATION = NUPHA,B36,130;%%
%\cite{AlvarezGaume:1981hn}

  L.~Alvarez-Gaume, D.~Z.~Freedman and S.~Mukhi,
  ``The Background Field Method And The Ultraviolet Structure Of The
  Supersymmetric Nonlinear Sigma Model,''
  Annals Phys.\  {\bf 134} (1981) 85.
  %%CITATION = APNYA,134,85;%%
 \bibitem{BCZ}
  %\cite{Braaten:1985is}
  E.~Braaten, T.~L.~Curtright and C.~K.~Zachos,
  ``Torsion And Geometrostasis In Nonlinear Sigma Models,''
  Nucl.\ Phys.\  B {\bf 260} (1985) 630.
  %%CITATION = NUPHA,B260,630;%%

  S.~Mukhi,
  ``The Geometric Background Field Method, Renormalization And The Wess-Zumino
  Term In Nonlinear Sigma Models,''
  Nucl.\ Phys.\  B {\bf 264} (1986) 640.
  %%CITATION = NUPHA,B264,640;%%


\bibitem{Polyakov1981}
  A.~M.~Polyakov,
  ``Quantum geometry of bosonic strings,''
  Phys.\ Lett.\  B {\bf 103}, 207 (1981).
  %%CITATION = PHLTA,B103,207;%%

%\cite{Callan:1985ia}
\bibitem{Callan:1985ia}
  C.~G.~Callan, E.~J.~Martinec, M.~J.~Perry and D.~Friedan,
  ``Strings In Background Fields,''
  Nucl.\ Phys.\  B {\bf 262}, 593 (1985).
  %%CITATION = NUPHA,B262,593;%%

%\cite{Giveon:1994fu}
\bibitem{Giveon:1994fu}
  A.~Giveon, M.~Porrati and E.~Rabinovici,
  ``Target space duality in string theory,''
  Phys.\ Rept.\  {\bf 244} (1994) 77
  [arXiv:hep-th/9401139].
  %%CITATION = PRPLC,244,77;%%



%\cite{Callan:1986jb}
\bibitem{Callan:1986jb}
  C.~G.~.~Callan, I.~R.~Klebanov and M.~J.~Perry,
  ``String Theory Effective Actions,''
  Nucl.\ Phys.\  B {\bf 278}, 78 (1986).

\bibitem{CallanThor}
  C.~G.~.~Callan and L. Thorlacius,
  ``Sigma Models in String Theory,''
  in Particles, Strings and Supernovae, Eds. A. Jevicki and C. I. Tan, World Scientific (1988).



\bibitem{MyersOdd}
  N. Kaloper and R. C. Myers,
  ``The O(dd) story of massive supergravity,''
  [arXiv:hep-th/9901045].
  %%CITATION = HEP-TH 9901045;%%

\bibitem{new}
C.~M.~Hull and R.~A.~Reid-Edwards,
  ``Gauge Symmetry, T-Duality and Doubled Geometry,''
  arXiv:0711.4818 [hep-th].
  %%CITATION = ARXIV:0711.4818;%%
G.~Dall'Agata, N.~Prezas, H.~Samtleben and M.~Trigiante,
  %``Gauged Supergravities from Twisted Doubled Tori and Non-Geometric String
  %Backgrounds,''
  arXiv:0712.1026 [hep-th].
  %%CITATION = ARXIV:0712.1026;%%




\end{thebibliography}
\end{document}